\documentclass[sn-mathphys,Numbered]{sn-jnl}

\usepackage{graphicx}%
\usepackage{multirow}%
\usepackage{amsmath,amssymb,amsfonts}%
\usepackage{amsthm}%
\usepackage{mathrsfs}%
\usepackage{physics}
\usepackage{lmodern}

\usepackage[title]{appendix}%
\usepackage{xcolor}%
\usepackage{textcomp}%
\usepackage{manyfoot}%
\usepackage{booktabs}%
\usepackage{algorithm}%
\usepackage{algorithmicx}%
\usepackage{algpseudocode}%
\usepackage{listings}%

\renewcommand{\Im}{\imaginary}

\begin{document}

\title[]{Chiral optical forces in a slot waveguide for separation of molecules in gas}

\author[1]{\fnm{Josep} \sur{Martínez-Romeu}}

\author[1]{\fnm{Daniel} \sur{Arenas-Ortega}}
\author[1]{\fnm{Iago} \sur{Diez}}



\author[1]{\fnm{Alejandro} \sur{Martínez}} \email{amartinez@ntc.upv.es}

\affil[1]{\orgdiv{Nanophotonics Technology Center}, \orgname{Universitat Politècnica de València}, \orgaddress{\street{Camino de Vera, s/n Building 8F}, \postcode{46022}, \city{Valencia}, \country{Spain}}}






\abstract{\textbf{}Chiral optical forces present an exciting avenue into the separation of enantiomers in an all-optical fashion. In this work, we explore via numerical simulations and analytical calculations the feasibility of the separation of chiral molecules by using guided light in a dielectric slot waveguide. Our results suggest that it is possible to separate [6]helicene enantiomers suspended in gas within several hours when applying optical powers of 100 mW.}





\maketitle   


Chirality is the property of objects that are not superimposable to their mirrored selves. The study of objects that satisfy this property is of great importance in different branches of science and technology. Of particular importance is the case of molecular chirality, which is related to the well-known phenomena of optical activity and circular dichroism, which allows us to distinguish between molecules exhibiting opposite chirality (called enantiomers). Remarkably, while two enantiomers have the exact same composition, they may show different chemical and optical properties\cite{Barron_2004}. Chiral medicines serve as an example, where one enantiomer might produce the desired pharmacological effect, whereas the other enantiomer might be toxic \cite{Smith2009}. Therefore, the efficient separation of chiral molecules is critical in diverse areas such as medicine, pharmaceutical and agricultural industries. Currently, there are several technologies to separate enantiomers, such as chiral high-performance liquid chromatography (HPLC), where the chiral compounds flow in a liquid or gas phase in contact with a chiral solid phase, which attracts one enantiomer more than the other, thus separating the initial mixture \cite{Okamoto2008}. All current chiral separation techniques face relevant problems, such as different chiral molecules needing different chiral agents to be separated and high costs for using and discarding the solvents in which separation occurs.

There has been recent interest in studying separation methods that take advantage of the forces that light exerts on enantiomeric entities \cite{Hayat2015,Genet2022} to overcome the limitations mentioned above of current technology. Interestingly, when light carries helicity, whether transversal or longitudinal, such optical forces can be chiral, in the sense that the same optical field can move each enantiomer toward opposite directions, thus producing separation. Many recent works have reported numerical calculations or theoretical studies \cite{Schnoering2021,Mun2020,Kravets2019,Cameron_2023,Golat2023} that show the feasibility of enantiomeric separation. Moreover, several experiments have shown how optical chiral forces can move enantiomers in different directions or at different velocities \cite{Jacques1981,Tkachenko2014,Yang2009}. Among the different approaches for separation, the use of dielectric waveguides seems particularly interesting. In such systems, optical chiral forces can be sustained over long distances while being transversally confined in regions of the order of half a wavelength, thus maximizing the optical intensity that interacts with the chiral entities \cite{Iago2024,Josep2024}.

In a previous study \cite{Josep2024}, it was shown that chiral pressure forces exerted by a quasi-circular polarized optical field guided along a strip dielectric waveguide could enable the separation of nanoparticles with small chirality immersed in water. This optical field is achieved by combining the fundamental TE and TM modes propagating with a 90º-phase shift, and shows longitudinal helicity, which is responsible for the chiral pressure force. While successful for nanoparticle separation, applying this technique to molecules in a liquid environment encounters a critical limitation: the induced forces are too weak to achieve separation within practical time frames using reasonable optical power.

Here, we studied the case of molecules ([6]helicene) flowing in air and analyzed the possibility of enantiomeric sorting by calculating the optical forces and performing particle tracking and time evolution simulations. In particular, we considered a slot waveguide design capable of supporting guided light with local helicity whilst enhancing the local forces in the slot region.

 
We calculated the optical forces exerted by the guided field on [6]helicene molecules by modeling the molecules as dipoles with electric, magnetic, and chiral polarizability ($\alpha_{\rm e}$, $\alpha_{\rm m}$, $\alpha_{\rm c}$, respectively). The force expression can be written as \cite{Golat2023}:

\begin{eqnarray}\label{eq:all_forces}
    {\bf F}=&\underbrace{{\bf \nabla}(\mathfrak{R}\alpha_\text{e}W_\text{e}\!+\!\mathfrak{R}\alpha_\text{m}W_\text{m}\!+\!\mathfrak{R}\alpha_\text{c}\omega\mathfrak{G}\! )}_\text{gradient force}
    \\&\underbrace{+2\omega(\mathfrak{I}\alpha_\text{e}{\bf p}_\text{e}\!+\!\mathfrak{I}\alpha_\text{m}{\bf p}_\text{m}\!+\!\mathfrak{I}\alpha_\text{c}\mathfrak{R}{\bf p}_\text{c}\! )}_\text{radiation pressure force}
    \\&\underbrace{-(\sigma_\text{rec}\mathfrak{R}{\bf \varPi}\!+\!\sigma_{\text{im}}\mathfrak{I}{\bf \varPi})/c
    -\omega(\gamma^\text{e}_\text{rec} {\bf S}_\text{e}\!+\!\gamma^\text{m}_\text{rec} {\bf S}_\text{m})}_\text{dipole recoil force}.\!\!\!
\end{eqnarray}

\noindent where $\omega$ is the angular frequency, $k$ is the wavenumber, $W_{\rm e} = \frac{1}{4}\varepsilon |{\bf E}|^2$ and
$W_{\rm m} = \frac{1}{4}\mu |{\bf H}|^2$ are the electric and magnetic energy densities, respectively, measured in $ \left[\mathrm{J}/ \mathrm{m}^3\right]$ units.
The helicity density is $\mathfrak{G} = \frac{1}{2\omega c} \mathfrak{I} \left( {\bf E} \cdot {\bf H}^* \right)$ $ \left[\mathrm{J}\cdot \mathrm{s}/\mathrm{m}^3\right]$, whose sign indicates the handedness of the optical field. The following field properties ${\bf S}_{\rm e} = \frac{1}{4\omega} \mathfrak{I} \left( \varepsilon \ {\bf E}^*\, \times {\bf E} \right) $ $\left[\mathrm{J}\cdot \mathrm{s}/\mathrm{m}^3\right]$ and $
{\bf S}_{\rm m} = \frac{1}{4\omega} \mathfrak{I} \left(\mu \, {\bf H}^*\times {\bf H} \right) $ $\left[\mathrm{J}\cdot \mathrm{s}/\mathrm{m}^3\right]$ yield respectively the electric and magnetic spin densities of the field. The complex Poynting vector is represented by ${\bf \varPi} = \frac{1}{2}{\bf E} \times {\bf H}^* $ $\left[\mathrm{W}/\mathrm{m}^2\right]$. The electric, magnetic and chiral momentum density of the light field are respectively   
${\bf p}_\text{e}=\frac{1}{2c^2}\mathfrak{R}{\bf \varPi}-\frac{1}{2}{{\bf \nabla} \times}{{\bf S}_\text{e}}$, ${\bf p}_\text{m}=\frac{1}{2c^2}\mathfrak{R}{\bf \varPi}-\frac{1}{2}{{\bf \nabla} \times}{{\bf S}_\text{m}}$ and $\mathfrak{R}{\bf p}_\text{c}=k({\bf S}_\text{e}+{\bf S}_\text{m})-\frac{1}{2\omega c}{{\bf \nabla} \times}{\mathfrak{R}{\bf \varPi}}$. The parameters of the recoil force depend on the product of polarizabilities: $\sigma_\text{rec}=\frac{k^4}{6 \pi}[\mathfrak{R}(\alpha_{\text{e}}^* \alpha_{\text{m}})+\left|\alpha_{\text{c}}\right|^2]$, $\sigma_\text{im}=\frac{k^4}{6 \pi}\mathfrak{I}(\alpha_{\text{e}}^* \alpha_{\text{m}})$, $\gamma^\text{e}_\text{rec}=\frac{k^4}{3\pi }\mathfrak{R}(\alpha_\text{e}^\ast\alpha_\text{c})$, $\gamma^\text{m}_\text{rec}=\frac{k^4}{3\pi }\mathfrak{R}(\alpha_\text{m}^\ast\alpha_\text{c})$ \cite{Golat2023}.  

In the proposed waveguide system, pressure forces are the dominant forces in the optical enantioseparation. The reason is the following. This technique relies on applying a sustained optical force field to displace enantiomers linearly over time, effectively surpassing the diffusive displacement caused by Brownian motion, which scales with the square root of time \cite{Iago2024}. This can be guaranteed if the force field is uniform in direction and strength and constant over time. Despite the significant Brownian motion suffered by molecules, the consistent push of pressure forces along the waveguide length ensures a continuous displacement that eventually leads to enantioseparation \cite{Golat2023,Josep2024}. In contrast, gradient forces are confined to the evanescent decay region near the waveguide walls with non-uniform direction and strength, so the interaction with molecules is spatially and temporally inconsistent. Furthermore, gradient forces for this particular system act in the transversal plane, thus not contributing to a longitudinal sorting. In summary, the force expression can be approximated by only considering the pressure forces, which depend on the imaginary part of the polarizabilities of the molecule as follows:
\begin{equation}\label{eq:pressure_forces}
        {\bf F}\approx\underbrace{2\omega(\mathfrak{I}\alpha_\text{e}{\bf p}_\text{e}\!+\!\mathfrak{I}\alpha_\text{m}{\bf p}_\text{m})}_{\text{achiral}}\!+\underbrace{2\omega\!\;\mathfrak{I}\alpha_\text{c}\mathfrak{R}{\bf p}_\text{c}}_{\text{chiral}}\!
\end{equation} 

The electric, magnetic, and chiral polarizabilities of [6]helicene enantiomers, named P-enantiomer ($\Im \alpha_{\rm c}>0$) and M-enantiomer ($\Im \alpha_{\rm c}<0$), were obtained from experimental circular dichroism measurements registered in \cite{chiral_database}. We selected 332 nm as the operating wavelength for enantioseparation, as it corresponds to helicene's peak chiral polarizability, and silicon nitride (SiN) as the constituent material of the slot waveguide due to its transparency at that wavelength, and its established use in fabricating photonic integrated circuits.


        


The use of the chiral pressure force to perform chiral separation along the longitudinal direction requires the existence of a quasi-circularly polarized field being guided along the waveguide. Such guided field is obtained from the superposition of a TE and a 90$^\circ$-shifted TM mode that need to be degenerate at the operating wavelength \cite{Josep2024}. If such modes are non-degenerate, the helicity of the guided light would be periodically changing (from positive to negative going through zero) along the waveguide length, and consequently the longitudinal chiral pressure force too, thus averaging to zero chiral force. 
The degeneracy condition requires that $n_{\rm TE}=n_{\rm TM}$, being $n_{\rm TE} $ and $n_{\rm TM}$, the effective index of the TE and TM guided modes, respectively.

The slot waveguide \cite{Yang2009} design is composed of two SiN ($n=2.1187$) strip waveguides with rectangular cross-section (132 nm wide $\times$ 80 nm thick), separated by a 30 nm air gap (the slot), and placed on a SiO$_2$ ($n=1.4802$) substrate. The slot is closed by a top cladding of SiO$_2$. The chosen slot width ensures strong confinement of the electromagnetic energy within the slot, and can be accomplished using standard fabrication techniques. We used a commercial Finite Element Method (FEM) mode solver (FemSIM by Synopsys, see also \cite{Iago2024,Josep2024}), to obtain the electric and magnetic field of the modes which are needed for computing the force with Eq.~\ref{eq:all_forces}. The solver also yielded the effective index of the TE and TM guided modes, used to find which combination of waveguide thickness and width ensured the degeneracy condition at the operating wavelength. This was accomplished for SiN strips of the before mentioned dimensions.

\begin{figure}[htb!]
    \centering
    \includegraphics[width=0.8\textwidth]{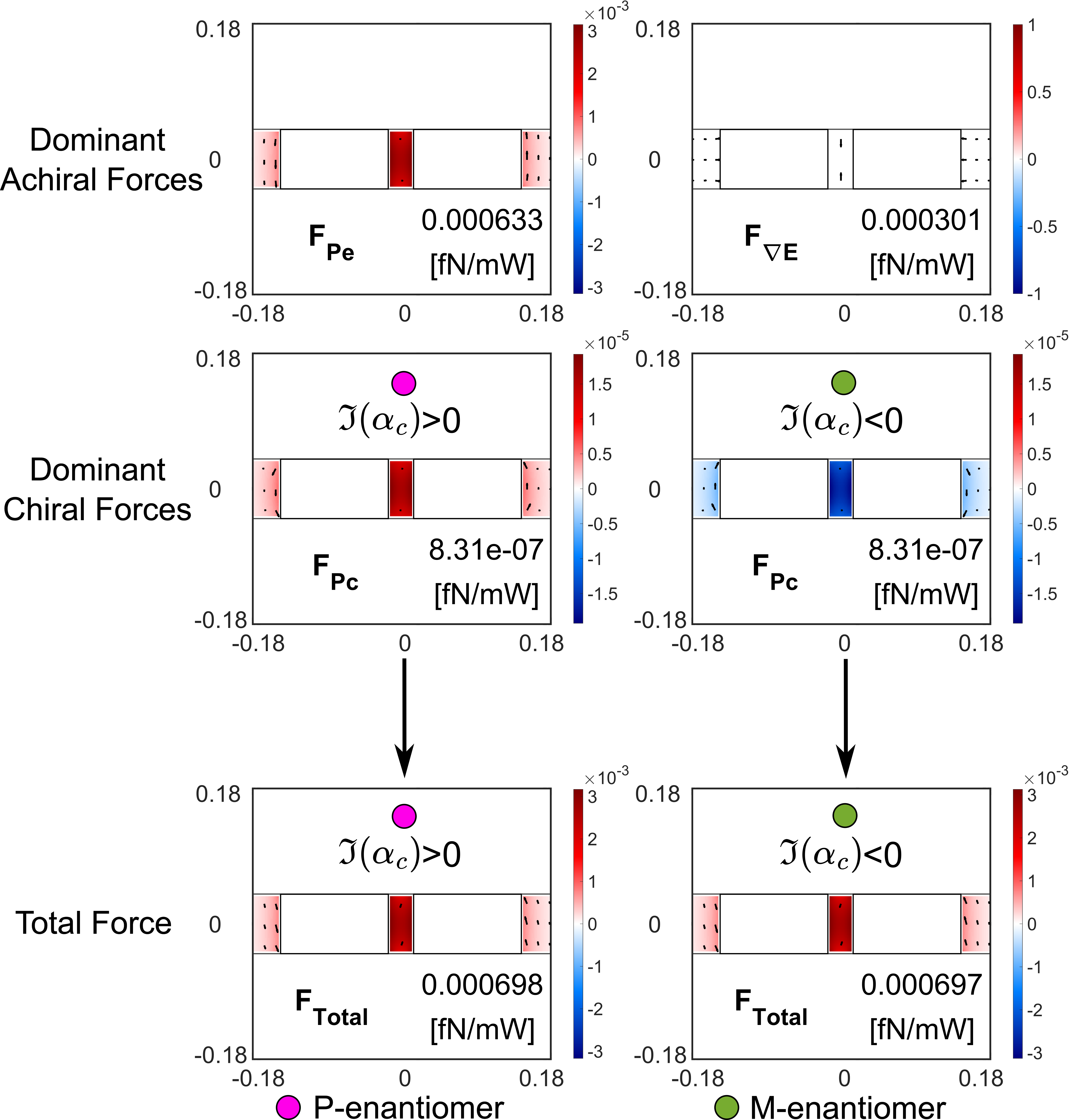}
    \caption{Optical forces exerted on [6]helicene enantiomers by the combined mode (TE plus 90$^\circ$-shifted TM) propagating along the SiN slot waveguide operating at a 332 nm wavelength and assuming a total power of 100 mW for the guided light. We represent the most dominant chiral force and achiral forces as well as the total force throughout the cross-section (XY plane) of the slot waveguide. The maximum value of the longitudinal achiral force on [6]helicenes is 3.15·$10^{-3}$ fN/mW , while the maximum longitudinal chiral force is 1.93·$10^{-5}$ fN/mW.}
    \label{ForcesSlot}
\end{figure}

The map of the dominant optical forces is plotted in Fig.~\ref{ForcesSlot}. Due to the chiral forces being opposite in direction for opposite enantiomers, there is a difference in the total force experienced by each enantiomer in the longitudinal direction. According to our results, the P-enantiomer experiences a longitudinal force of 0.003167 fN/mW, whereas the M-enantiomer is subjected to a force of 0.003129 fN/mW. This difference, though subtle, can enable the longitudinal separation. In the transversal plane, both enantiomers experience an achiral attractive (gradient) force towards the center of the slot. While this transversal force arising from gradient forces is stronger than the chiral forces, the molecules will not be trapped due to the Brownian motion. Therefore, by using longitudinal chiral forces, Brownian motion can be overcome, and the system can sort enantiomers. 
As mentioned before, this mechanism relies on the fact that while Brownian motion expands the enantiomer distribution width at a rate proportional to the square root of time (diffusion process), the optical pressure forces in a fluid move the distributions at a rate linearly proportional to time. This makes it possible to perform sorting even when the difference in the enantiomer velocities is small, as in this work with [6]helicene molecules. As further explained in \cite{Josep2024}, it is just a matter of waiting sufficient time until the desirable separation degree is achieved.

\begin{figure}[htb!]
    \centering
    \includegraphics[width=1\textwidth]{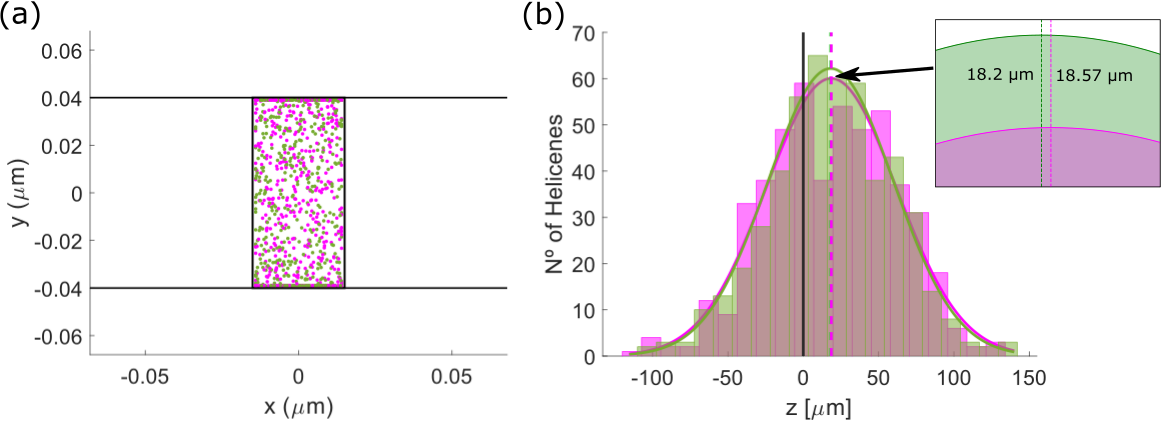}
    \caption{First step of the sorting simulation based on a particle tracking algorithm. Position of the enantiomers in the slot after 0.1 seconds of tracked time. (a) Cross-sectional view of the slot waveguide with the [6]helicenes (500 points per enantiomer). Slot dimensions: 80 nm high $\times$ 30 nm wide. (b) Longitudinal position histogram of [6]helicene enantiomers. The centers of the P- and M-enantiomer clouds (gaussian distributions) moved 18.57 $\mu$m and 18.20 $\mu$m respectively.} 
    \label{Particle_tracking_air}
\end{figure}

The sorting simulation was divided into two steps: a particle tracking simulation followed by a time evolution calculation. First, the optical force map of the system is fed to a tracking algorithm (explained in detail in \cite{Josep2024}). The algorithm solves iteratively the overdamped Langevin equation to track the position of both [6]helicene enantiomers along the volume of the slot waveguide. The tracking is done individually per enantiomer, and repeated 500 times to obtain statistically meaningful results. Although a two-dimensional electromagnetic simulation was performed to obtain the electric and magnetic fields of the modes, the force field is $z$-invariant due to the translational symmetry along the longitudinal axis. The initial position of the molecules was uniformly distributed throughout the slot cross-section (80 nm $\times$ 30 nm) and longitudinally from $z=-500$ nm to  $z= 500$ nm. 
The tracking simulation was performed during 0.1 seconds for [6]helicenes suspended in air and assuming 100 mW of power guided along the slot waveguide. We chose air as the carrier fluid due to its lower viscosity ($1.8$·$10^{-5}$ Pa·s) compared to water's ($10^{-3}$ Pa·s), which substantially reduces the sorting time. 

Results show that both enantiomers end up being uniformly distributed throughout the cross-section of the slot waveguide (Fig.~\ref{Particle_tracking_air}a). After 0.1 seconds, the 500 final $z$-positions per enantiomer form gaussian distributions, which we will refer to as enantiomer clouds (Fig.~\ref{Particle_tracking_air}b). The last average longitudinal position of each cloud is similar ($z_+=$18.57 $\mu$m and $z_-=$18.20 $\mu$m), revealing that 0.1 seconds is not sufficient time to separate the enantiomers in space. From the mean value between $z_+$ and $z_-$, we obtained the average achiral travelling speed: 183.60 $\mu$m/s. Since the chiral pressure force is 200 times weaker than the achiral pressure force, the sorting speed is very low. Because of the computational limitations, tracking for longer times is not possible. that is why we needed a second step in the sorting simulations.

\begin{figure}[ht!]
    \centering
    \includegraphics[width=1\textwidth]{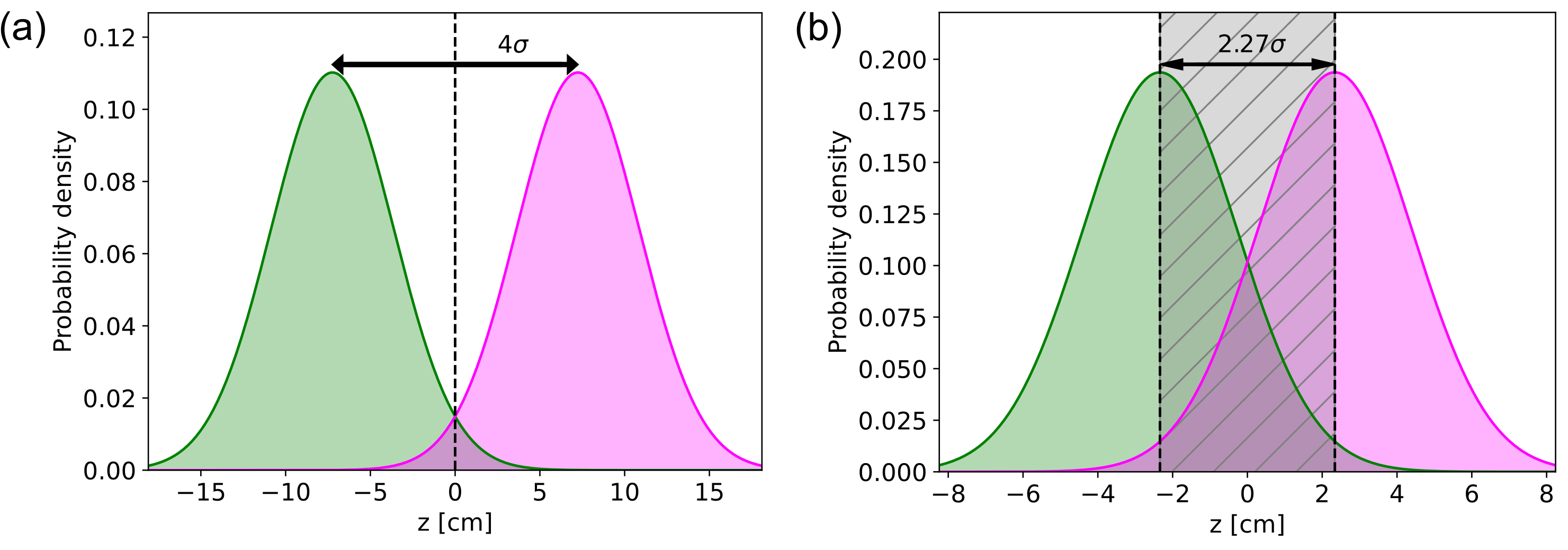}
    \caption{Second step of the sorting simulation based on a time evolution algorithm. Schematics for the two different separation methods. (a) 4$\sigma$ separation case, where we separate the mixture of molecules into two zones delimited by the dashed line. (b) 2.27$\sigma$ separation case, where we separate in three zones delimited by the dashed lines. In the left(right) zone, the M(P) enantiomers are present with an enantiomeric fraction of 97.72\%. In the middle zone, the mixture is not separated. Here, the displacement due to the achiral forces has been canceled out so that the mid position between both distributions is kept at $z=0$.} 
    \label{time_evolution_air}
\end{figure}

The second step of the sorting simulation consists of evolving the average position and width of the gaussian distributions in time. The initial conditions of this time evolution were the average positions and widths of the last iteration from the first step of the tracking simulation (Fig.~\ref{Particle_tracking_air}b). The velocity at which the two enantiomer distributions travel and spread is obtained from the average achiral force and ratio between maximum chiral and achiral force within the slot. Full details are described in \cite{Josep2024}. With this method, we obtained the time needed to separate both enantiomer clouds and achieve regions of space where the enantiomeric fraction of the amount of molecules is 97.72\% (see Suppl. Information). The strong achiral pressure force causes excessive cloud displacement, making this application impractical for photonic integrated circuits, which are limited to several tens of centimeters in size. To mitigate this effect, we propose inducing a fluid flow opposing the achiral force, with a flow velocity matching the achiral traveling speed. This method has been used in previous optical chromatography experiments \cite{Bacillus2006,Hart2007}. This fixes the $z$-position of the midpoint between the clouds (e.g. at middle of the waveguide length) while keeping the chiral force difference between enantiomers.
(Fig.~\ref{time_evolution_air}). We considered two scenarios in which 100\% of the injected amount of molecules are sorted, or only 50\% of the amount but quicker. These two conditions correspond to a distance of separation between the centers of the clouds of $z_{+}-z_{-} = 4\sigma_{z}$ and $z_{+}-z_{-} = 2.27\sigma_{z}$, respectively, where $\sigma_{z}$ is the width of the position distribution. The widths of both enantiomer distributions should be equal because of the same diffusion coefficient.




The resulting separation distance and time are summarized in Table \ref{tab:Separation}. While 100\% of the amount was separated in 20 hours (14.5 cm), 50\% took just 7 hours (4.67 cm). In water, the separation (100\%) would take 5.6 days (further explanation and data in water are provided in the Supplementary Material).




\begin{table}[htb!]
    \centering
    \begin{tabular}{c|ccc}
        Amount of separated molecules  & Separation distance & Time & EF \\ 
          \toprule
        100\% & 14.5 cm (4$\sigma$) & 20 hours & 97.72 \% \\ 
        50\% & 4.67 cm (2.27$\sigma$) & 7 hours & 97.72 \%
    \end{tabular}
    \caption{Summary table with separation results. Assuming the availability of 300 mm SiN wafers, the slot waveguides should be capable of $2.27\sigma$ and $4\sigma$ separation of [6]helicene.}
    \label{tab:Separation}
\end{table}    


In summary, our simulation results show that a dielectric slot waveguide might be a suitable system to produce chiral separation of [6]helicene enantiomers suspended in gas due to the strong confinement of the fields inside the slot. Using an optical power of 100 mW distributed to a combination of the TE and TM mode, separation can be carried out within a time-scale of 7 to 20 hours and total waveguide lengths of about 15 to 30 cm, thus fitting in commercial 300 mm SiN wafers \cite{Giewont2019}. While the fabrication methods constrain the separation feasibility of [6]helicene, this type of slot waveguide might provide a possible future use case for the separation of bigger chiral molecules, which would be subjected to stronger forces, thus enabling faster sorting in shorter waveguides.

Acknowledgments: The authors acknowledge funding from the European Commission under the CHIRALFORCE Pathfinder project (grant no. 101046961). A.M. acknowledges partial funding from the Conselleria de Educación, Universidades y Empleo under the NIRVANA Grant (PROMETEO Program, CIPROM/2022/14).

Data available in Zenodo repository at \cite{Data}.

\bibliography{sample}

\pagebreak 
\begin{center}
\textbf{\large Supplementary Information: Chiral optical forces in a slot waveguide for separation of molecules in gas}
\end{center}
\setcounter{equation}{0}
\setcounter{figure}{0}
\setcounter{table}{0}
\setcounter{page}{1}
\makeatletter
\renewcommand{\theequation}{S\arabic{equation}}
\renewcommand{\thefigure}{S\arabic{figure}}
\renewcommand{\theHfigure}{S\arabic{figure}}

\section{Polarizabilities of [6]helicene molecules}
\begin{figure}[htb!]
    \centering
    \includegraphics[width=1\textwidth]{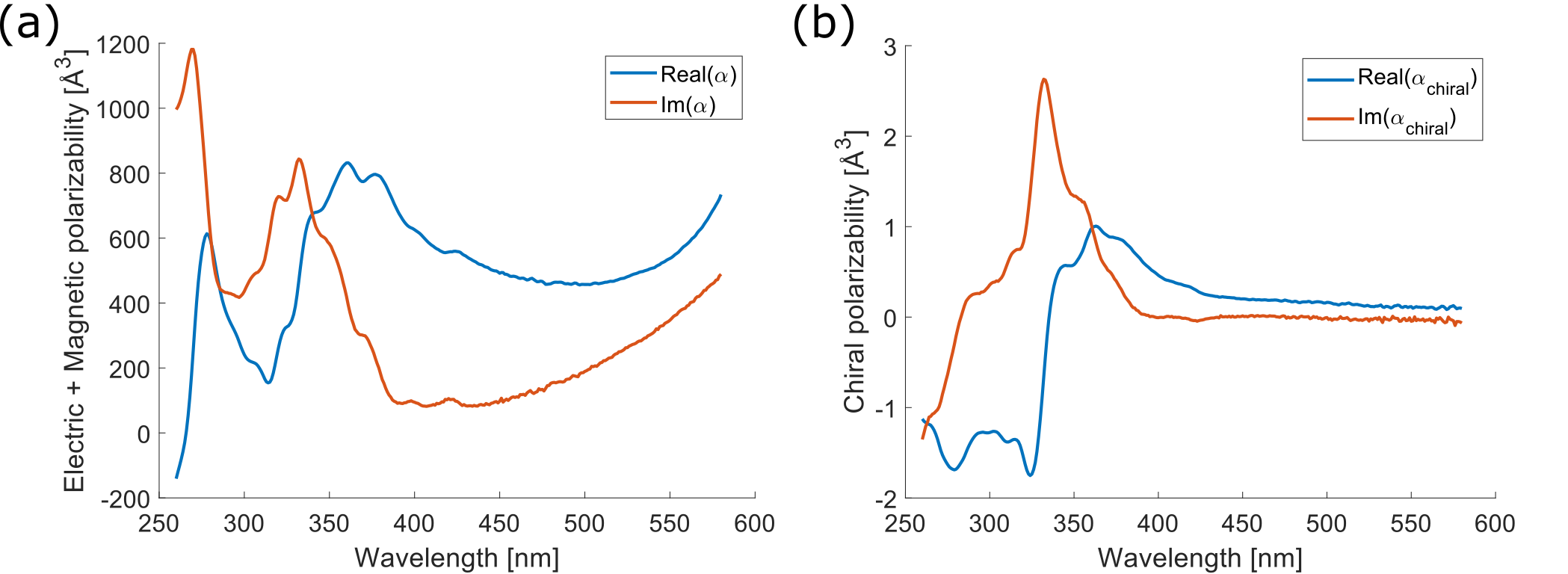}
    \caption{[6]Helicene achiral (a) and chiral (b) polarizabilities. Values retrieved from \cite{chiral_database}.
    }
    \label{Spectrum_Helicene}
\end{figure}

The values of the electric and magnetic polarizabilities are depicted in Fig. \ref{Spectrum_Helicene}a and the chiral polarizability in Fig.~\ref{Spectrum_Helicene}b. The imaginary part of the polarizabilities was obtained directly from circular dichroism experimental measurements \cite{chiral_database}: the electric and magnetic polarizabilities from the extinction spectra and the chiral polarizability from the circular dichroism signal. The real part was obtained by applying the Hilbert transform to the imaginary parts.  The individual contribution of the electric and magnetic polarizability to the extinction spectra cannot be known. Thus, we assume $\alpha_{\rm m} = 0$, so that the extinction is attributed exclusively to $\Im \alpha_{\rm e}$. Notice that the maximum of the imaginary part of the chiral polarizability of [6]helicene takes place at a wavelength of 332 nm, as seen in Fig.\ref{Spectrum_Helicene}b. 

\section{Separation of [6]helicene enantiomers in water}

\begin{figure}[htb!]
    \centering
    \includegraphics[width=0.8\textwidth]{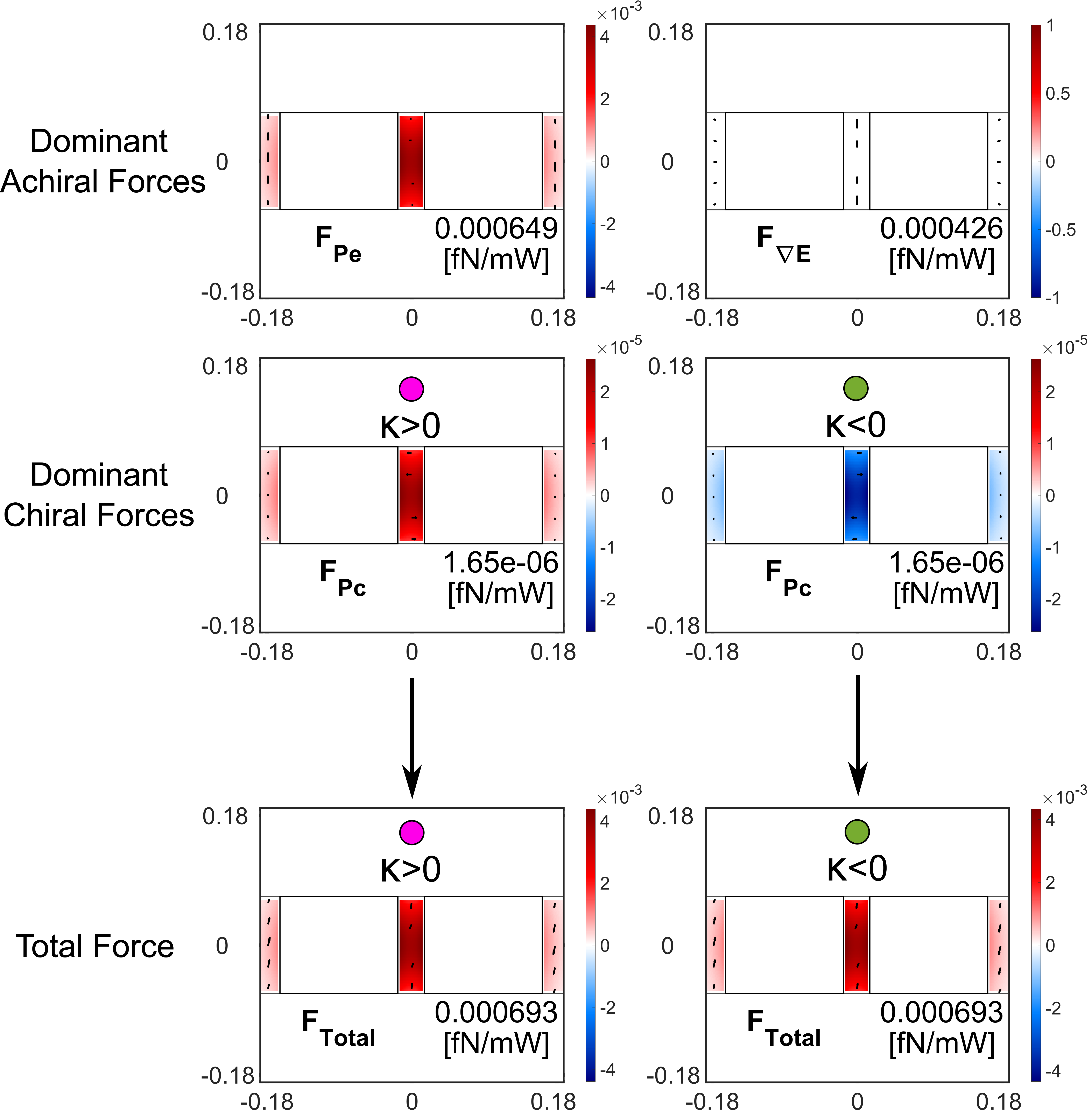}
    \caption{Optical forces acting on [6]helicene throughout the cross-section of the degenerate slot waveguide system. We show the dominant achiral, and dominant chiral forces as well as the total force. Maximum longitudinal achiral forces in water 0.004365 fN/mW, maximum chiral forces in water 0.000026 fN/mW.}
    \label{ForcesWater}
\end{figure}

The optical forces acting on [6]helicene throughout the cross-section of the degenerate slot waveguide in water are shown in Fig.~\ref{ForcesWater}. 
The positive enantiomer feels a longitudinal force of 0.004391 fN/mW while the negative enantiomer is subject to a force of 0.004339 fN/mW. 

\begin{figure}[ht!]
    \centering
    \includegraphics[width=\textwidth]{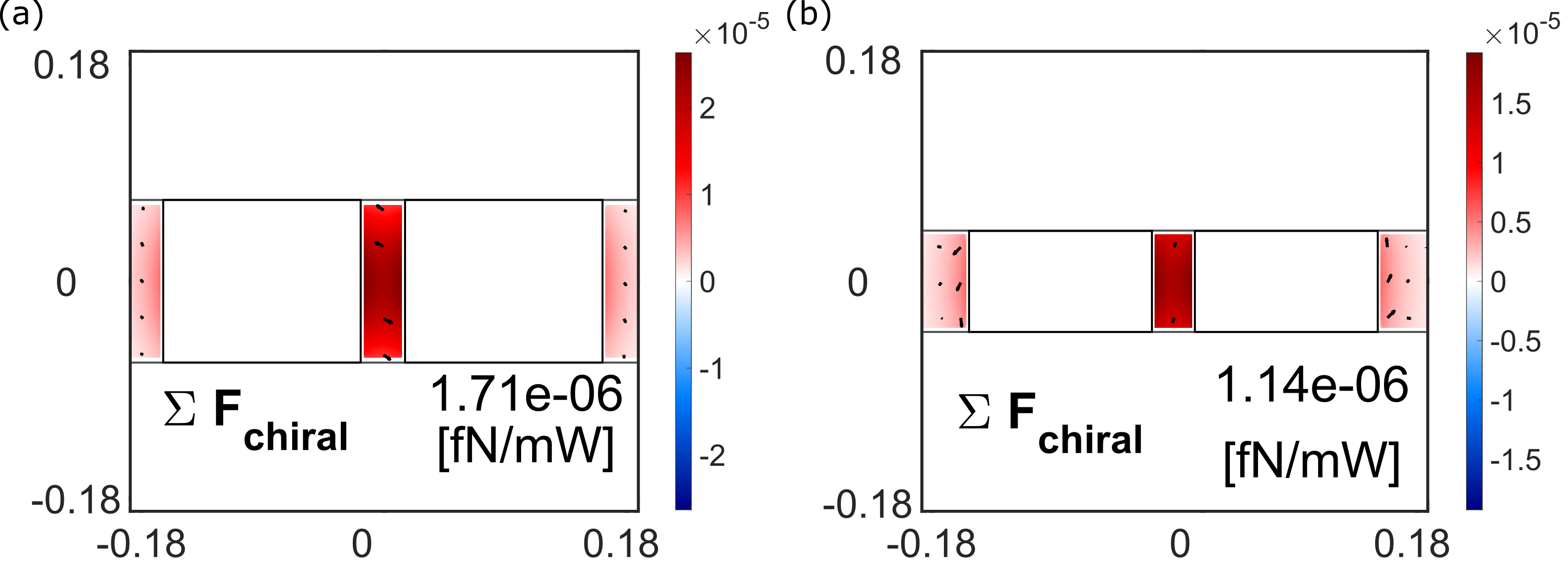}
    \caption{Comparison of the chiral forces between the slot waveguide in water and the slot waveguide in air.  (a) Water as medium. Maximum longitudinal force value is 0.000026 fN/mW. (b) Air as medium. Maximum longitudinal force value is 0.000019253 fN/mW.} 
    \label{Force_comparison}
\end{figure}

 The difference in forces between having water or air in the slot waveguide, is shown in Fig.~\ref{Force_comparison}: stronger forces occur in water. In principle, this is opposite to what one would expect to happen in a slot waveguide. In such system, the field intensity in the slot is proportional to $n^{2}_{\rm SiN}/n^{2}_{\rm slot}$. The intensity of the field should be stronger when the permitivity is smaller in the slot, so the field intensity, and consequently the forces, should be stronger in the case of air. This does not occur due to the difference in waveguide dimensions that are needed to maintain the degeneracy of the modes. In air the required dimensions are 132 nm thickness $\times$ 80 nm width, whereas in water are 138 nm thickness $\times$ 125 nm width. 
 
 Although stronger forces occur in the water slot waveguide, the velocity that the enantiomers acquire due to the optical field is much higher for the air slot system due to air's lower viscosity. The velocity of spherical particles of radius R is given by:
\begin{equation}
    v=\frac{F}{6\pi \eta R}
\end{equation}
\noindent where $\eta$ is the viscosity of the fluid ($1.8$·$10^{-5}$ Pa·s for air, $10^{-3}$ Pa·s for water) and $F$ the optical force. This explains why enantioseparation occurs faster in air than in water. For comparison, the achiral traveling speed in air was 183.60 $\mu$m/s, and in water was 9.2 $\mu$m/s. 

\begin{figure}[htb!]
    \centering
    \includegraphics[width=\textwidth]{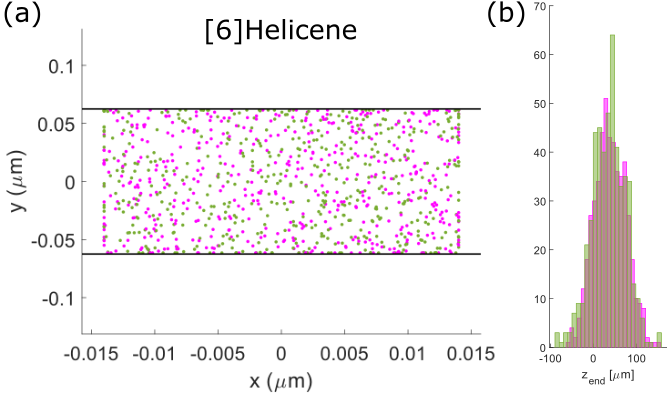}
    \caption{Particle tracking of [6]helicenes situated inside the gap of the slot waveguide during 4 seconds. The forces were calculated for 100 mW power.  The [6]helicenes enantiomers end up uniformly distributed throughout the cross-section. In the longitudinal direction, the enantiomers mainly move due to the achiral pressure force. At this time, the centers of the clouds for each enantiomer are 38.7 $\mu$m and 34.9 $\mu$m. Thus, the clouds travel at an average achiral speed of 9.2 $\mu$m/s.}
    \label{Particle_tracking_water}
\end{figure}

The results of the particle tracking are shown in Fig.~\ref{Particle_tracking_water}. 
The expected sorting time (for an enantiomeric fraction of 97.72\%) was obtained with the time evolution algorithm. For $z_{+}-z{-} = 4\sigma_{z}$ (100 \% of molecules separated), the sorting time is 5.6 days and the separation distance is 5.28 cm. For $z_{+}-z{-} = 2.27\sigma_{z}$ (50\% of molecules separated), the sorting time would be 3 days and 12 hours, and the separation distance would be 2.36 cm. The traveled distance due to the achiral force would still be excessive (beyond tens of centimeters), so the compensation with an opposite flow rate is needed for water environment too.



\newpage

\section{Enantiomeric fraction definition} 

A number to evaluate the purity of a separated racemic mixture of enantiomers is the enantiomeric fraction, which is defined as:

\begin{equation}
    (+)\text{-EF}=\frac{N_{+}}{N_{+}+N_{-}}
\end{equation} 

\begin{equation}
    (-)\text{-EF}=\frac{N_{-}}{N_{+}+N_{-}}
\end{equation} 

Where $N_{-}$, $N_{+}$ are the negative (M) and positive (P) enantiomers in a given spatial region. When the enantiomeric fraction in a given region is superior to 95 \% the mixture has been sufficiently separated. 

     
\end{document}